\documentclass[11pt,twoside]{article}
\usepackage[dvips]{graphics}
\usepackage{Cargesepasp}
\newcommand{\oversim}[2]{\protect{\mbox{\lower0.5ex\vbox{%
   \baselineskip=0pt\lineskip=0.2ex
   \ialign{$\mathsurround=0pt #1\hfil##\hfil$\crcr#2\crcr\sim\crcr}}}}} 
\newcommand{\simgreat}{\mbox{$\,\mathrel{\mathpalette\oversim>}\,$}} 
\newcommand{\simless} {\mbox{$\,\mathrel{\mathpalette\oversim<}\,$}} 
\def\edcomment#1{\iffalse\marginpar{\raggedright\sl#1\/}\else\relax\fi}
\reversemarginpar

\begin{document}
\title{The Birth, Evolution and Death of Star Clusters}
 \author{Pavel Kroupa} 

\affil{Institut f\"ur Theoretische Astrophysik\\
Tiergartenstr.~15, D-69121 Heidelberg, Germany}

\begin{abstract}
A dense-enough gas-accumulation evolves, over a few~Myr of
intensifying star formation, to an embedded cluster.  If it contains a
sufficient amount of mass, O~stars form and explosively expel the
remaining gas, whereas poorer clusters reduce their embryonic gas
content more gradually. The sudden expulsion of gas unbinds most of a
rich cluster, but a significant fraction of it can condense by
two-body interactions to become an open cluster despite a
star-formation efficiency as low as 30~per cent.  Poorer clusters
survive their gradual mass loss more easily, but have short,
relaxation-limited life-times. Pleiades-like clusters may thus form as
nuclei of expanding OB associations, by filling their tidal radii and
having large (1--2~pc) core-radii. A 'main-sequence' of clusters is
thus established. Ultimately, a cluster dies an explosive death
through the ever shortening relaxation time, and leaves a remnant that
consists of about 4--10 stars arranged in a highly hierarchical and
thus long-lived system. Dynamical mass segregation in very young
clusters is extremely rapid, and heats a cluster substantially, which
is partially off-set by the cooling from the disruption of primordial
binaries.
\end{abstract}


\section{Introduction}
\noindent
Star clusters fascinate because some are beautifully evident to the
unaided eye, and because their birth, life and death remain
mysterious. Wonderful examples 'magically' lying together in one
quadrant on the summer sky (southern hemisphere) are M42 (the Orion
Nebula Cluster, ONC), and the Pleiades and Hyades Clusters. Their ages
are $\tau_{\rm cl}\approx1$~Myr, 100~Myr and 600~Myr, respectively,
and latest research indicates they may form approximately one
evolutionary sequence. Such clusters may be the origin of a
significant proportion of Galactic-field (GF) stars, which is one
reason why we want to understand their behaviour.  In this
contribution (KII), the aim is to convey some theoretical aspects
concerning the birth, evolution and death of open clusters, but I
stress that much remains to be worked out in this exciting field.
Complementary texts are Kroupa (2000a, KI; 2000b, KIII).

A few quantities useful for roughly assessing the global state of a
cluster are the half-mass diameter crossing-time [Myr], 

\hfill $t_{\rm cr} = 4.2\,\left(M_{\rm st}/\left(\epsilon\,100\,M_\odot\right)
\right)^{-1/2} \left(R_{0.5}/1\,{\rm pc}\right)^{3/2}$, \hfill (1)

\noindent 
the median relaxation-time [Myr] for a purely stellar system (Binney \&
Tremaine 1987, BT)

\hfill$t_{\rm rel} = \left(21/ \ln(0.4\,N)\right) \left(M_{\rm st} /
100\,M_\odot\right)^{1/2} \left(\overline{m}/1\,M_\odot \right)^{-1}
\left(R_{0.5} / 1 \, {\rm pc}\right)^{3/2}$, \hfill (2)

\noindent
and the three-dimensional velocity dispersion [pc/Myr] in the cluster,

\hfill $\sigma = 0.47\,\left(M_{\rm
st}/\left(\epsilon\,100\,M_\odot\right)\right)^{1/2}
\left(R_{0.5}/1\,{\rm pc} \right)^{-1/2}$. \hfill (3)

\noindent Here, $R_{0.5}$ is the half-mass radius, $M_{\rm
st}/\epsilon = M_{\rm g}+M_{\rm st}$, $M_{\rm g}$ being the gas mass
and $\epsilon$ the star-formation (sf) efficiency, and $M_{\rm
st}=N\,\overline{m}$ is the mass of the cluster containing $N$ stars
that have an average mass $\overline{m} = 0.36 \, M_\odot$, assuming
the universal (GF) IMF (Kroupa 2000c) with stellar masses in the range
$0.01-50\,M_\odot$, $G=4.49\times10^{-3}\,{\rm pc}^3/(M_\odot\,{\rm
Myr}^2)$, and 1~pc/Myr~$\approx$~1~km/s.

For one massive star with $m=10\,(20)\,M_\odot$, the GF IMF implies
$N_{10 \left(20\right)}=371\,(1155)$ stars with
$m\in\left[0.01,10\right]\, \left(\left[0.01,20\right]\right)
\,M_\odot$. These numbers are useful for determining if an embedded
cluster is likely to contain stars massive enough to remove the natal
gas within a time $\tau_{\rm g}<t_{\rm cr}$, i.e. 'explosively'. In
what follows, {\it poor} clusters contain $N\simless10^3$ stars,
whereas {\it rich} clusters contain $N\simgreat10^3$~stars.

Typical numbers for a few well-known clusters are $\tau_{\rm
cl}\simless1$~Myr, $N = 5000-10000, R_{0.5}\approx0.45$~pc (ONC),
$\tau_{\rm cl}\simless1$~Myr, $N\approx200, R_{0.5}\approx0.5$~pc
($\rho$~Oph, still embedded in gas), $\tau_{\rm cl}\approx100$~Myr,
$N\approx3000, R_{0.5}\approx3$~pc (Pleiades), and $\tau_{\rm
cl}\approx600$~Myr, $N\approx2000, R_{0.5}\approx4$~pc (Hyades).

\section{Birth and Morphology}
\label{sec:morph}
The birth of a cluster undoubtedly requires some relatively large gas
mass ($M_{\rm g}\simgreat500\,M_\odot$) to be squeezed into a
relatively small volume ($R_{0.5}\simless 0.5$~pc). This is evident
from many observational results (e.g. contributions in Lada \& Kylafis
1999; Clarke, Bonnell \& Hillenbrand 2000; these proceedings), but how
a molecular cloud decides to do this remains to be understood
theoretically. Calculations using the SPH-approximation to treat
gas-dynamics during the earliest stages of the formation of a cluster
(prior to stellar feedback) shows that it is heavily sub-structured
(Klessen \& Burkert 2000), which is also evident from observations
(e.g. Kaas \& Bontemps 2000). Observational evidence indicates that sf
begins slowly and accelerates (Palla \& Stahler 2000), probably owing
to the increasing density and probably additional sf being triggered
by feedback, with a characteristic duration, $t_{\rm sf}$, of a
few--many~Myr.  The star-gas system probably contracts due to
dynamical friction of the stars on the gas (Saiyadpour, Deiss \& Kegel
1997), and/or the stars that 'freeze-out' of the gas fall towards the
cluster centre if the turbulent length-scale is smaller than the
region from which the stars condense (ie. small proto-star--proto-star
velocity dispersion). As the proto-stars continue forming, their
orbits in the cluster rapidly (within $t_{\rm cr}$) virialise, so that
the stellar system should always be close to virial equilibrium in the
potential that is determined mostly by the mass-dominating gas.  As a
result of these processes, the sub-structure washes out on the global
$t_{\rm cr}$ time-scale. In subregions, in which $t_{\rm cr,sub} <
\tau_{\rm sf}$, where $\tau_{\rm sf}\approx0.1$~Myr is the time for
the formation of an individual proto-star (e.g. Wuchterl \&
Tscharnuter 2000), and $t_{\rm cr,sub} \ll t_{\rm cr}$, the forming
proto-stellar clumps collide to form more massive clumps that are more
likely to collide, and can accrete even more radially in-flowing gas
with low specific angular momentum, causing contraction of the
subregion which speeds-up this process. It is thus in the localised
density maxima that massive stars are expected to form through
collisional accumulation (Bonnell, Bate \& Zinnecker 1998; Stahler,
Palla \& Ho 2000), which is more or less supported by observations
(Megeath et al. 1996), but the growth of massive stars through
accretion alone remains an important possibility (e.g. Norberg \&
Maeder 2000).

Termination of the sf process occurs abruptly through the devastating
action of one or more O~stars in rich embedded clusters. When the
first O~star 'ignites', an ultra-compact~HII region (UCHII, radius
$R\simless0.1$~pc) develops which is stable and 'long-lived' (a
few~$10^5$~yr), being confined by the large pressure of the
star-forming gas, before finally breaking out to become a compact HII
region (CHII, $R\in\left(0.1-0.3\right)$~pc) with complex gas
structures, and finally an extended HII region (EHII,
$R\simgreat$~few~pc) (Garcia-Segura \& Franco 1996). The evolution
from the UCHII to an EHII region occurs rapidly, the heated ionised
gas expanding with at least the sound velocity (about 12~km/s), but
also being driven by the fast wind (a few~100~km/s, e.g. Lamers, Snow
\& Lindholm 1995) emanating from the hot surface of an O~star. Even
prior to the eruption of the UCHII, large quantities of the
surrounding gas are removed through the massive outflows expelled from
the region containing the massive stars (Churchwell 1997; 1999).
Star-formation ceases more gradually by the removal of unused gas
through less-massive outflows and winds powered by lower-mass stars in
poor embedded clusters (Matzner \& McKee 2000).

Observational evidence (section~6.2 in Matzner \& McKee 2000 for an
over\-view) suggests that $\epsilon\simless0.4$ in embedded clusters.
Matzner \& McKee (2000) theoretically estimate that $0.30 \simless
\epsilon_{\rm MM} \simless 0.50$ for poor clusters, $\epsilon$
essentially being determined by the rate with which sf can continue as
the increasing number of outflows collect and discard increasing
amounts of unused gas.  For rich embedded clusters,
$\epsilon<\epsilon_{\rm MM}$, because once the O~star 'ignites',
essentially all sf is terminated throughout the cluster, no transition
time existing during which sf dies down.

The ONC is in the CHII$\rightarrow$EHII phase, whereas $\rho$~Oph is
still in the embedded phase but may never have O~stars. The Pleiades
and Hyades have evolved well beyond the EHII stage, and how they
survived explosive gas expulsion was an unsolved problem until very
recently.

\section{Cluster Survival and Ultimate Death}
\label{sec:clsurv}
Once an embedded cluster forms, three mass-loss mechanisms work over
different time-scales towards unbinding it: (i) Expulsion of
embryonic gas (approximately during first 0--5~Myr), (ii) mass loss
from evolving stars (significant after about 3~Myr), and (iii)
stellar-dynamical evaporation and ejections of stars (all times).

\vskip 1mm
\noindent {\bf Gas-expulsion:} It has been realised since a long time
that the expulsion of significant amounts of gas from an embryonic
cluster has serious implications for its survival.  Hills (1980),
Mathieu (1983), and Elmegreen (1983) provide ground-breaking
analytical results, and the first $N$-body experiments
($N\le100$~stars) were performed by Lada, Margulis \& Dearborn
(1984). These pioneers arrived at the general result that if
gas-expulsion occurred instantly (i.e. $\tau_{\rm g}<t_{\rm cr}$
essentially, so that the stellar orbits cannot adjust to the rapidly
varying potential), as is the case in clusters with O~stars, and if
$\epsilon<0.5$ then an unbound association results, although the
$N$-body results indicated that $\epsilon\approx0.4$ still allows a
small part of the embedded cluster to 'hang-on' as an expanded but
bound entity. A contraction of the stellar system relative to the gas
leads to higher effective $\epsilon$, allowing more of the embedded
cluster to survive after gas expulsion. Further variations on the
theme incorporate an initial cold collapse, in which however only a
small fraction of the gas needs to be removed to lead to an unbound
association, unless the removal occurs just at the beginning or after
violent relaxation (and thus after a new, contracted virial equilibrium
of the stellar system is achieved with a much higher effective
$\epsilon$).  

The particularly noteworthy result emerged that open clusters, such as
the Pleiades, could not have formed with O~stars {\it and}
$\epsilon\simless0.5$. Since $\epsilon>0.5$ was never observed, the
implication is that Galactic clusters can only form with stellar IMFs
that are truncated near $m\approx5-10\,M_\odot$. In this respect, the
future fate of the ONC, which contains OB~stars that clearly rapidly
drove out the unused gas some time ago, is interesting, in as much as
it has a velocity dispersion that is too large for virial equilibrium
of the stellar system (e.g. Hillenbrand \& Hartmann 1998).

In clusters without~O~stars, the gas is expelled over a time
comparable to the duration of the most intense sf period, i.e.
$\tau_{\rm g}\approx t_{\rm sf}$ (a few~Myr), since the stars that go
'on-line' immediately add their outflows to the general erosive
commotion. The above-mentioned pioneering work established that under
these conditions, the stellar system expands, but because it has
enough time to adjust to the varying potential, roughly half of it can
relax to form a bound cluster filling its tidal radius.

More recent numerical experiments (e.g. Goodwin 1997; Geyer \& Burkert
2000) confirm the above results, and Adams (2000; also these
proceedings) analytically estimates the fraction of stars that have
velocities {\it below} the escape velocity after gas and the unbound
stars escape, assuming, however, that the stellar system is more
concentrated than the gas, implying an effective $\epsilon\approx0.9$, as
noted by Geyer \& Burkert (2000). It is thus not surprising that Adams
finds that substantial clusters form despite an assumed $\epsilon<0.4$
(overall for gas + stars).  The above results were obtained under
simplifying assumptions, such as neglecting the Galactic tidal field
and mass loss from evolving stars, both of which further increase the
critical $\epsilon$ required for a bound Galactic cluster, probably
well above $\epsilon=0.4$, so that the formation of a Pleiades-like
cluster continued to remain a mystery.

All of the research summarised above assumed that the stellar system
can be treated as collision-less, i.e. that near-neighbour encounters
can be neglected while the gas is being expelled. The assumption
appears reasonable, but a detailed look uncovers that this is
flawed. Very recently, Aarseth's (1999) code {\sc Nbody6}, which
treats close stellar encounters accurately and computationally
efficiently using special mathematical transformations, has been
augmented by a rapidly varying back-ground potential (Kroupa, Aarseth
\& Hurley 2000, KAH).  The resulting code {\sc GasEx} has already been
used in a study of the future evolution of the ONC under the
assumption that $\epsilon=0.3$, and that it was in virial equilibrium
for 0.6~Myr (to model the UCHII phase) when the O~stars removed the
surplus gas explosively. A local Galactic tidal field, and
state-of-the art stellar evolution (Hurley, Pols \& Tout 2000) is
treated, and the (surprising) result is that a Pleiades-like cluster
{\it readily} condenses from the radially expanding flow. About $1/3$
of the initial number of stars condense, which is substantially more
than implied under similar conditions by the above mentioned research.

Thus, the Pleiades most probably formed from an ONC-like object, and
the mechanism for this must be two-body encounters that allow density
differences to grow as the system expands, causing a part of the
radial flow to be re-directed into orbital motions about the centre of
expansion thus forming a substantial bound cluster. Clearly, this is
an exciting field of research, and details still have to be worked
out, but the results available so far suggest that Galactic clusters
form with large core radii ($\approx1-2$~pc) and filling their tidal
radii, as a result of the expansion after gas loss. Binary disruption
is very efficient during the embedded epoch, and the surviving binaries
are always hard when the cluster reaches a relaxed state in the
Galactic tidal field, which may be called a {\it cluster main
sequence}, since global evolution thereafter depends only on the
number of stars within the tidal radius (fig.~1 in K3), assuming
everything else (IMF, primordial binaries) is universal. Any
initial mass segregation, imposed either as a result of cluster
formation or through rapid dynamical mass segregation (see below),
will be much reduced after the expansion, but some memory of it
remains.  A new epoch of mass segregation begins as soon as the
'cluster main sequence' is reached (KAH).  A notable example of a very
young and very massive open/globular cluster, that is just at the
stage of having expanded to fill it's tidal radius, may be Cygnus~OB2
(Kn\"odlseder 2000). Each Galactic cluster should thus be associated
with an expanding population of co-eval stars which amounts to about
$2/3$ of the total number of stars formed in the one event, and these
extended moving groups will be identifiable with the upcoming
astrometric satellite missions DIVA (R\"oser 1999) and GAIA (Gilmore
et al. 1998).

Finally, the Pleiades will appear similar to the Hyades when 600~Myr
old (Portegies Zwart et al. 2000).

\vskip 1mm
\noindent{\bf Mass-loss from evolving stars:}
Stars with $m\simgreat8\,M_\odot$ explode as supernovae (sn), their
remnants having $1-2\,M_\odot$, but usually being lost from the
cluster owing to the sn kick ($>10$~km/s, e.g. Portegies Zwart,
Kouwenhoven \& Reynolds 1997). The last sn ($\approx8\,M_\odot$ star)
occurs at an age of about 40~Myr. Thereafter mass-loss occurs via
planetary nebulae leaving white dwarf remnants with masses in the
range $0.5-1.2\,M_\odot$ (Weidemann 1990).

The ratio of the mass lost from a cluster to the initial stellar mass
is, for the universal GF IMF (Kroupa 2000c, eq.~2),

\hfill$\Delta M/M_{\rm st, init} = 
0.733\,m_{\rm to}^{-0.3} - 0.169\,m_{\rm to}^{-1.3} - 0.215$, \hfill (4)

\noindent
assuming $m\ge8\,M_\odot$ stars kick out their remnants, and that for
less massive stars the remnants have a mass of about $1\,M_\odot$ and
are retained in the cluster. Thus, within 40~Myr a cluster looses
about 17~per cent of its mass (sn explosions), within 100~Myr
(turn-off mass $m_{\rm to}\approx5\,M_\odot$) about 23~per cent is
lost, within 600~Myr ($m_{\rm to}\approx2\,M_\odot$) 31~per cent is
lost, and within about 14~Gyr ($m_{\rm to}\approx0.9\,M_\odot$) a
cluster would loose in total about~35~per cent of its initial mass
owing to stellar evolution alone (stellar evolution times are from
Hurley, Pols \& Tout 2000).  A cluster thus suffers under the
significant mass loss from evolving stars (e.g. de la Fuente Marcos
1997).  However, less mass is lost over progressively longer
time-intervals, so that it ultimately becomes negligible compared to
dynamical 'evaporation'.

\vskip 1mm
\noindent{\bf Stellar-dynamical evaporation and ejection:} Stars in a
'main-sequence cluster' suffer many weak, long-distance 'encounters'
with other cluster stars, so that the kinetic and potential energies
are constantly being re-arranged among cluster members ({\it two-body
relaxation}, e.g Lee \& Goodman 1995). Energy equipartition in the
cluster potential leads to the less massive stars gaining kinetic
energy and the heavier ones gaining potential energy, thus sinking
towards the cluster centre. After many such two-body encounters, a
low-mass star may find itself with a positive energy relative to the
cluster. It is lost from the cluster, typically after one Galactic
orbit usually by exiting through one of the Lagrange points
(e.g. Portegies Zwart et al. 2000), but on rare occasions it may
suffer an encounter while traversing the cluster and may be scattered
back into membership.  Estimates based on simplified model clusters
(single stars of equal mass) suggest that a cluster evaporates within

\hfill $t_{\rm life}\approx 20\,t_{\rm rel}$\hfill (5)

\noindent
(e.g. BT, section~8.4), but this estimate breaks down for a young
cluster that does not fill it's tidal radius. This is evident in
models that have the same $N$ but different $R_{0.5}$, and thus very
different initial $t_{\rm rel}$, leading to the same $t_{\rm life}$
(Kroupa 1995b, K3). The above equation may be used only when the
cluster is relaxed in the tidal field. Thus, a poor cluster with
typically $N\approx200$ after gas expulsion and $R_{0.5}\approx2$~pc
has $t_{\rm life}\approx0.6$~Gyr, whereas for the Pleiades $t_{\rm
life}\approx3$~Gyr.

Rare, close encounters between single stars can lead to the ejection
of stars with relatively large velocities at the expense of the more
massive stars that gain binding energy in the cluster. These processes
are, however, much rarer than the loss through evaporation (BT and
references therein). However, the large abundance of primordial
binaries in realistic clusters increases the ejection rate and
ejection velocities substantially. The models of K3 show that about
$4-5$ times as many stars are ejected with velocities $v_{\rm
ej}\simgreat5$~km/s in clusters with primordial binaries than in
clusters without binaries. The models also demonstrate that initially
more concentrated clusters loose a larger percentage of their stars
through energetic ejection events. Binary-star encounters are not very
significant for expelling stars as a cluster ages (also pointed out by
de la Fuente Marcos 1997), because the surviving binaries are hard
with small interaction cross sections (see above).  Overall, this
channel for loosing stars from a cluster amounts to less than 10~per
cent of the initial stellar population of the 'main-sequence cluster'
(K3), and since a cluster ultimately dissolves completely,
evaporation, driven by cluster expansion through mass loss from
evolving stars, remains the dominant mechanism for stellar loss.  It
is clear though that the velocity distribution of ejected stars
contains information about the typical birth configuration, for
example of massive stars (see KIII).

Owing to the preferred loss of the least massive and single stars, a
Galactic cluster approaches an idealised state, being composed
predominantly of stars with similar masses around $m_{\rm to}$ and an
enhanced proportion of hard binaries. At late evolutionary times, the
binary proportion can achieve levels as high a 80~per cent (K3), even
if the primordial proportion was only $1/3$ (de la Fuente Marcos
1997).

\vskip 1mm
\noindent{\bf Death:}
As a consequence of mass loss, the tidal radius, 

\hfill $R_{\rm tide}\approx \left(M_{\rm st}(t)/\left(3\,M_{\rm gal}
\right)\right)^{1/3} \,R_{\rm GC}$, \hfill (6)

\noindent
where $M_{\rm gal}\approx5\times10^{10}\,M_\odot$ is the Galactic mass
within the solar radius $R_{\rm GC}\approx 8.5$~kpc, contracts, the
density within remaining approximately constant. However, $R_{0.5}$
need not decrease monotonically as a consequence. It may increase
intermittently if a core forms that contracts thus heating the rest of
the cluster. This is the case for rich clusters, but details on the
evolution along the putative cluster main sequence remain to be worked
out.  The contracting $R_{\rm tide}$ implies a decrease of $t_{\rm
rel}$, so that the cluster's dynamical evolution speeds up.  Open
clusters thus die, in a sense, explosively, their population dwindling
at an ever increasing rate.  This can be seen in the evolutionary
curves in fig.~1 of K3, as well as in the models of de la Fuente
Marcos (1997).  The very final stage in a cluster's life sees the
formation of a stable few-body system, the {\it cluster remnant}. Such
a remnant consists of only a few ($4\simless N_{\rm rem}\simless 10$)
stars that are arranged in a stable, strongly hierarchical multiple
system (de la Fuente Marcos 1997; de la Fuente Marcos
1998). Individual tight binaries in it need not be primordial.  Such
an object can be very long-lived and difficult to find, but a complete
census of all strongly hierarchical systems would give insights into
the number of Galactic clusters ever born in our neighbourhood.

An 'unnatural' but not unlikely death of a star cluster can occur if
it encounters a molecular cloud (Theuns 1992).

\section{Primordial Binaries: Cooling}
\label{sec:hac}
Changing to a more fundamental topic: in order to assess the role of
realistic primordial binary systems as heating and cooling sources,
and to study early mass segregation, a library of clusters has been
generated (table~1 in KI). These have $N=800, 30000, 10000$, the same
initial central ONC density, the same $t_{\rm cr}$, assume different
primordial binary populations, but no initial mass segregation. These
models are set-up in theoretical virial equilibrium (Aarseth, H\'enon
\& Wielen 1974). Here focus is on the models with Taurus--Auriga-like
primordial binary-orbital parameters of Kroupa (1995a, K2) having a
primordial binary proportion $f_{\rm tot}=1$.

\begin{figure}
\plotfiddle{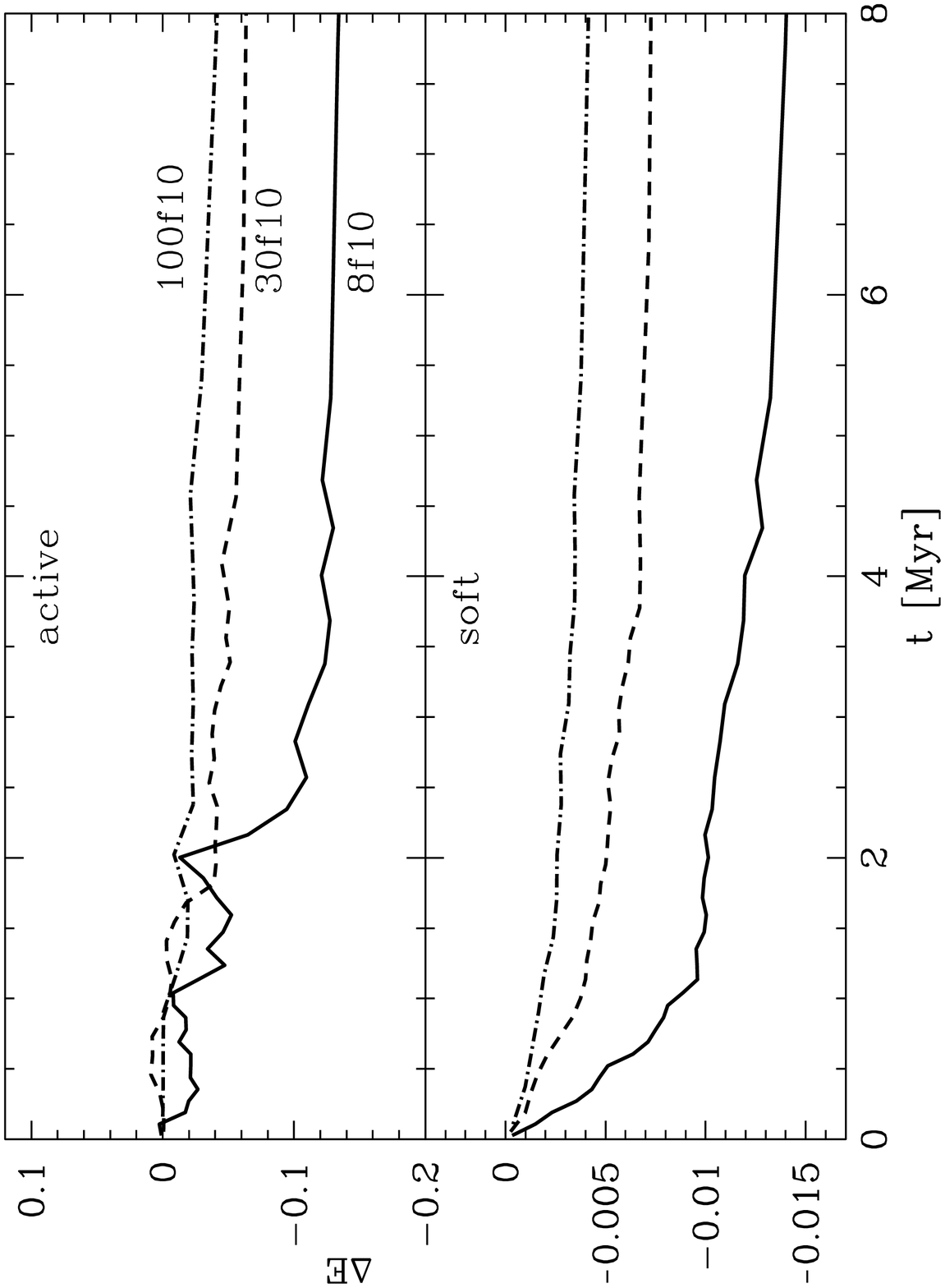}{6.8cm}{-90}{40}{40}{-160}{230}
\caption{\small Evolution of the total binding energy in 'active' and
'soft' binaries, scaled to the initial cluster kinetic energy.  Models
are as in KI and KIII, and the data are averages of $N_{\rm run}$
computations.}
\label{fig:ebins}
\end{figure}

As discussed in KI and KIII, soft binaries are typically disrupted on
a crossing-time scale (fig.~2 in KI), whereas hard ones harden thereby
injecting energy into the cluster which consequently heats up and
expands.  Disruption of binaries requires energy, cooling a
cluster. How significant cooling and binary heating is for very young
clusters can be estimated from the total binding energy in 'soft'
binaries, $E_{\rm bs} = \sum \left|e_{\rm bs,i}\right|$, such that
$\left|e_{\rm bs,i}\right|<e_{\rm k}$, $e_{\rm
k}=0.5\,\overline{m}\sigma_{\rm 3D}^2$ being the average kinetic
energy, with $\overline{m}$ the average stellar mass and $\sigma_{\rm
3D}$ the three-dimensional velocity dispersion. 'Active' binaries have
$E_{\rm ba} = \sum \left|e_{\rm ba,i}\right|$, such that $e_{\rm k}
\le \left|e_{\rm ba,i}\right| \le 100\,e_{\rm k}$. The evolution of
$\Delta E_{\rm s} \equiv \left(E_{\rm bs}(0)-E_{\rm
bs}(t)\right)/E_{\rm k}(0)$ and $\Delta E_{\rm a} \equiv \left(E_{\rm
ba}(0)-E_{\rm ba}(t)\right)/E_{\rm k}(0)$, where $E_{\rm k}(0) =
0.5\,N\,\overline{m}\sigma_{\rm 3D}$ is the approximate total kinetic
energy content of the cluster, is shown in Fig.~\ref{fig:ebins}.

Thus, $\Delta E(t)<0$ (i.e. {\it cluster cooling}) in all cases except
for active binaries which heat for very brief times. The energy input
into the cluster is negligible though. What is interesting (and
somewhat surprising) is that the active binaries are efficient cooling
agents, especially for clusters with $N\simless 1000$. For these, the
energy 'soaked up' by the binaries amounts to about 12~per cent of the
initial kinetic energy. Soft binaries take $\simless 0.01\,E_{\rm k}$
out of the field, and are thus negligible cooling agents. Models with
a log-normal GF period distribution and $f_{\rm tot}=0.6$ always have
smaller $\Delta E$ than the models shown here that assume the 'K2 binary
population'. The cooling of the clusters is the reason why the central
density remains higher than in clusters containing initially fewer or
no binaries (fig.~4 in KI; fig.~2 in Kroupa, Petr \& McCaughrean
1999). 

The above is a preliminary account of a 'dash in this direction', and
more details concerning the energy transfer from active to hard
binaries (i.e. hardening of 'actives' to $\left|e_{\rm
b}\right|>100\,e_{\rm k}$) need to be worked out. That hard binaries
can have a substantial influence on the long-term evolution of a
cluster by being energy sources that counteract core-collapse in
massive clusters has been realised for a long time (e.g. Giannone \&
Molteni 1985; Giersz \& Spurzem 2000).

\section{Mass Segregation}
\begin{figure}
\plotfiddle{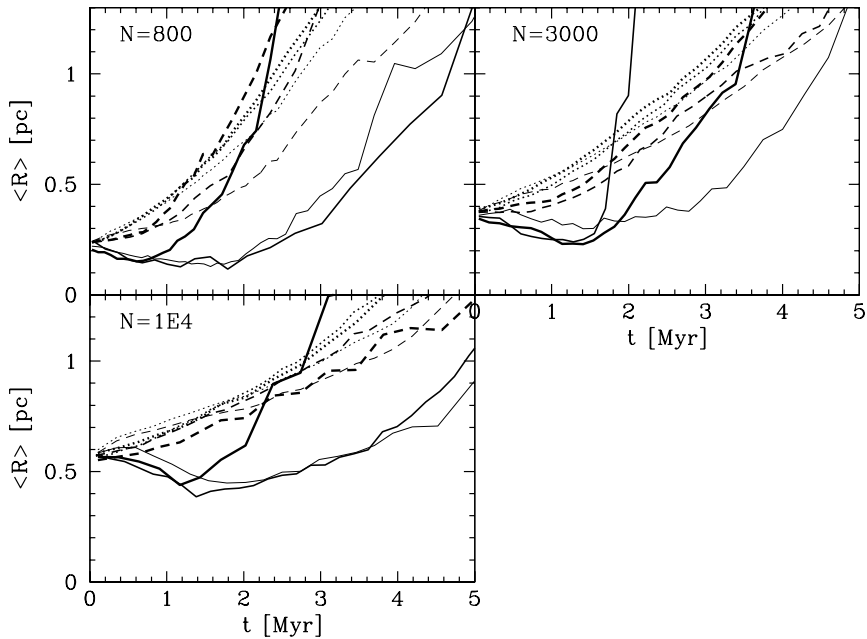}{7.4cm}{-90}{45}{45}{-180}{250}
\caption{\small The average radius of stars with $m\ge8\,M_\odot$
(solid lines), $1\le m < 8\,M_\odot$ (dashed lines) and BDs ($0.01\le
m\le 0.08\,M_\odot$, dotted lines). In decreasing curve thickness:
$f_{\rm tot}=1, 0.6, 0$ (same models as in
Section~\ref{sec:hac}). Table~1 in KI is augmented here by the models
100f10, 100f06 and 100f00 (lower left panel) which have the same GF
IMF (Kroupa 2000c) as the other models displayed here. }
\label{fig:mrad}
\end{figure}

Massive systems sink towards the centre thereby gaining potential
energy which {\it heats} the cluster. This is a consequence of {\it
energy equipartition}, as described above. The consequences of this
are seen in fig.~4 in KI, which shows that the central density
decreases for a wide range of very young clusters. That this is not
due to binary-star heating alone is evident from the expansion of the
single-star clusters. The observational consequences of this are an
expanding halo of BDs and M~dwarfs (Fig.~\ref{fig:mrad}).

The time-scale for {\it mass-segregation} to complete is not very well
known, this being an active area of research (e.g. Bonnell \& Davies
1998, who however employed a collision-less code to a collisional
problem), especially because of the need to understand trapezium-type
sub-systems in star clusters (Mermilliod 2000), and the associated
implications for the formation mechanisms of massive stars (Bonnell et
al. 1998).  An analytical estimate for the time-scale for
mass-segregation (more precisely, for the equipartition time-scale)
was obtained by Spitzer (1987, p.74),

\hfill $t_{\rm equ}\approx \left(\overline{m}/m_{\rm h}\right)t_{\rm
rel}$, \hfill(7)

\noindent where $\overline{m}, m_{\rm h}$ are the average and
heavy-star mass, respectively. Thus, for the examples shown in
Fig.~\ref{fig:mrad} ($\overline{m}=0.36\,M_\odot, m_{\rm
h}\approx20\,M_\odot$) one obtains $t_{\rm equ}=0.02\,t_{\rm
rel}<0.3$~Myr {\it in all cases}.  However, this estimate breaks down
as soon as the density structure of the cluster has changed
significantly, with further evolution slowing. Fig.~\ref{fig:mrad}
suggests that mass segregation completes by about~1~Myr in all
models. Note also that only the massive stellar sub-system contracts.

\begin{figure}
\plotfiddle{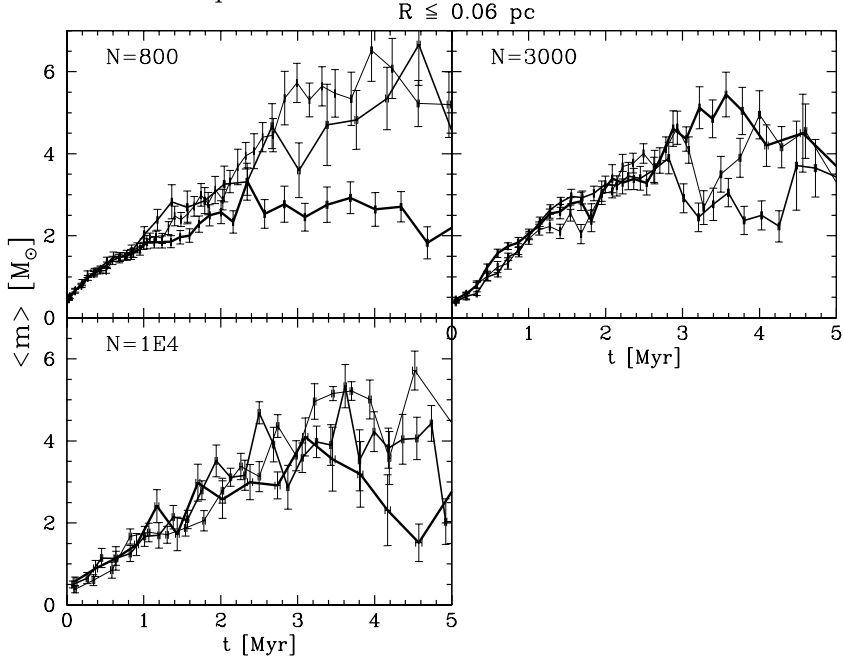}{7.4cm}{-90}{45}{45}{-180}{250}
\caption{\small The average stellar mass within $R\le0.06$~pc (in
decreasing thickness: $f_{\rm tot}=1, 0.6, 0$ for the same models as
in Fig.~\ref{fig:mrad}).}
\label{fig:mmass}
\end{figure}

The increase with time of the average stellar mass near the cluster
centre is shown in Fig.~\ref{fig:mmass}. Note that the increase is
linear until the core disrupts (massive stars are ejected), as is
evident in Fig.~\ref{fig:mrad}. The slope of $\overline{m}(t)$ depends
on the IMF (here Salpeter, eq.~2 in Kroupa 2000c), and additional
computations show a less rapid increase for a steeper IMF for massive
stars. In the present models, the average mass in the centre of the
cluster increases by a factor of about~10 within 3~Myr.

\vskip 1mm
\noindent {\bf Heating:} that the cluster is heated through mass
segregation is evident in figs.~4, and~5 in KI by a decreasing central
density and increasing $R_{0.5}$, and by increasing average radii
(Fig.~\ref{fig:mrad}), even for clusters without primordial binaries.
A very rough estimate of the significance of this heating source can
be attempted by comparing the change in binding energy of the massive
stars, $\Delta E_{\rm b,m}=G\,M_{\rm m}^2/\left(4\,R_{\rm
m}^2\right)$, where $R_{\rm m}$ is the average radius of $m\ge
8\,M_\odot$ stars at maximum contraction (Fig.~\ref{fig:mrad}), to the
cluster binding energy, $E_{\rm cl}=-G\,M_{\rm st}^2/\left(2\,R_{\rm
st}\right)$, where $R_{\rm st}$ is the initial average radius of all
stars (Fig.~\ref{fig:mrad}), 

\hfill $\Delta E_{\rm b,m}/E_{\rm cl} = \left(M_{\rm m}/M_{\rm
st}\right)^2 \, R_{\rm st}/\left(2\,R_{\rm m}^2\right)$, \hfill (8)

\noindent where $M_{\rm m}/M_{\rm st}=0.17$ (Kroupa 2000c).  One
obtains $\Delta E_{\rm b,m}/E_{\rm cl} = 0.14$, 0.22, 0.22 ($N=800,
f_{\rm tot}=1.0, 0.6, 0$), 0.11, 0.09, 0.06 ($N=3000, f_{\rm tot}=1.0,
0.6, 0$), and 0.04, 0.05, 0.04 ($N=10^4, f_{\rm tot}=1.0, 0.6, 0$).

Heating through initial mass segregation is thus a larger contribution
to the energy budget of a cluster than is cooling through the
disruption of active binaries for $N\simless 1000$, but comparable for
$N\simgreat 3000$. 

\vskip 1mm
\noindent {\bf Massive sub-system:} Once the massive stars assemble
near the cluster core, they more or less decouple from the rest of the
cluster (Hills 1975), because their velocity dispersion is
significantly smaller than that of the less massive stars, which pass
through the core too rapidly to interact significantly. The massive
stars interact, exchanging companions for more massive contemporaries,
ejecting the divorced partners, and being ejected themselves when one
of the involved systems hardens sufficiently
(Fig.~\ref{fig:mrad}). Under rare circumstances the stars may collide
and form even more massive rejuvenated exotic stars which may continue
interacting in the core or briefly roam Galactic space if ejected, as
studied in detail by Portegies Zwart et al. (1999).

The relaxation time of the core with radius $R_{\rm c}$ containing
$N_{\rm c}$ massive stars, $t_{\rm rel,c}\approx0.1\,\left(N_{\rm
c}/{\rm ln}N_{\rm c}\right)\,t_{\rm cr,c}$ (BT), so that the core will
relax, or significantly change it's energy, within about one crossing
time, $t_{\rm cr,c}$, if $N_{\rm c}\simless50$.  In the core, massive
stars pair-up to binaries, if they are not already with massive
partners. Such a binary can have a significant impact on the energy of
the core if it's binding energy, $e_{\rm
b}=-G\,\overline{m}^2/\left(2\,a_{\rm int}\right)$, is comparable to
the binding energy of a massive star to the core, $E_{\rm b,m}\approx
-G\,N_{\rm c}\overline{m}^2/\left(2\,R_{\rm c}\right)$, where
$\overline{m}$ is the average stellar mass in the core, {\it and} if
the cross section for encounters is sufficiently large. The former
condition can be written $\gamma\equiv E_{\rm b,m}/e_{\rm b} =
\left(2\,N_{\rm c}/\beta\right)^{1\over2}$, and the latter requires
the binary-star semi-major axis to be comparable to the cross-section
for one encounter per $\beta$~crossing times in the core. From the
core surface density, one obtains $a_{\rm int}\approx\sqrt{2}R_{\rm
c}/\left(N_{\rm c}\,\beta\right)^{1\over2}$.  Thus, $\gamma\simless10$
if $N_{\rm c}\simless 50$, the massive stars being in the core for
$\beta\approx1\,t_{\rm cr,c}$. The ejection of massive stars is very
efficient under these conditions since an encounter is likely to
liberate an amount of kinetic energy comparable to $-e_{\rm b}$.  On
the other hand, a cluster that contains $N_{\rm c}\simgreat 100$ OB
stars is less efficient in ejecting it's members ($\gamma\simgreat
10$).

The life-time of such a decoupled core is thus short. This can also be
seen in Fig.~\ref{fig:mrad}, where $N_{\rm c}\simless37$ for all
models.  Thus, the ``Trapezium'' ($N_{\rm c}\approx5$, each with a
mass of $\approx20\,M_\odot$, and $R_{\rm c}\approx0.05$~pc, giving
$t_{\rm rel,c}\approx 0.017$~Myr) in the core of the ONC is probably
in the final stage of it's decay.

\vskip 1mm
\noindent{\bf Ejection of massive stars:} The decay of the core
implies the ejection of OB stars, which is dealt with in more detail
in KIII.

\vskip 3mm 
\noindent

{\small This work was supported by DFG grant KR1635, and made use of
Aarseth's (1999) {\sc Nbody6} code.



\end{document}